\documentclass[11pt,a4paper]{article}

\usepackage{amsfonts,amsmath,amssymb}
\usepackage{mathrsfs}
\usepackage{amscd}
\usepackage{color}
\usepackage{graphicx}

\font\tengot=eufm10  \font\sevengot=eufm7  \font\fivegot=eufm5

\newfam\gotfam
\textfont\gotfam=\tengot
\scriptfont\gotfam=\sevengot
\scriptscriptfont\gotfam=\fivegot

\begin{document}
	
	
	
\title {Gravitational fields of axially symmetric compact objects in 5D space-time-matter gravity}
	

\author{J.L. Hern\'andez--Pastora\thanks{ Universidad de Salamanca, https://ror.org/02f40zc51.  e-mail address: jlhp@usal.es. ORCID:orcig.org/0000-0002-3958-6083}\ \\
Departamento de Matem\'atica Aplicada. Facultad de Ciencias,  and  \\
Instituto Universitario de F\'\i sica Fundamental y Matem\'aticas.\\
  Universidad de Salamanca. Spain }

\date{\today}

\maketitle

\vspace{-10mm}

\begin{abstract}
	
	In the standard Einstein's theory the exterior gravitational field of any  static and axially symmetric stellar object can be  described by means of a single function from which we  obtain  a metric into a four-dimensional space-time. In this work we present a generalization of those  so called Weyl solutions to  a  space-time-matter metric in a five-dimensional manifold within  a non-compactified Kaluza-Klein theory of gravity. The arising field equations reduce to those of vacuum Einstein's gravity when the metric function associated to the fifth dimension is considered to be constant.  	The calculation of the geodesics allows to identify the existence or not of different behaviours of test particles, in orbits on a constant plane,  between the two metrics.
	
	In addition, static solutions on the hypersurface orthogonal to the added dimension but with time dependence in the five-dimensional metric  are also obtained. The consequences on the variation of the rest mass, if the fifth dimension is identified with it, are studied.

 \end{abstract}

PACS numbers:  04.50.-h, 04.50.Cd, 04.50.Kd

Additional PACS No(s).: 04.20.Cv, 04.20.-q 

\large


\section{Introduction}

Kaluza\cite{kaluza} demonstrated that five-dimensional General Relativity (GR) contains both Einstein’s four-dimensional theory of gravity and Maxwell’s theory of electromagnetism. Few years later Klein \cite{klein} suggested the compactification of the fifth dimension as a plausible physical basis for the restriction on the coordinates. Kaluza-Klein theory unified not only  gravity and electromagnetism but also  matter and geometry for the photon  appeared in four dimensions as a manifestation of an  empty five-dimensional space-time. That interpretation of gravity   as pure geometry assumes the same paradigm used by  Einstein, and in addition the Kaluza mechanism is a minimal extension  of GR in the sense that there is no modification  to the mathematical structure of Einstein's theory just only changing the tensors indices running over one more value.

The question that one immediately asks oneself is why has no fifth dimension been observed in nature. Kaluza avoided the question simply demanding that all derivatives with respect to the fifth dimension vanish, or in other words by assuming that the physics takes place on a four-dimensional hypersurface into the five-dimensional universe (that is the so called {\it cylinder condition}). 
Consequently, another question arising is  why  does physics appear to be independent of the additional coordinates?
It was firstly proposed that extra dimension does not appear in physics because it is compactified and unobservable on experimentally accesible energy scales. This approach was  a dominant paradigm in higher-dimensional unification theories \cite{compactifing}. Another way to handle with the extra dimensions is to regard them  as mathematical artifacts of a more complicated underlying theories but which do not correspond to physical coordinates, as for example by replacing the affine geometry  of Einstein's relativity with projective geometry \cite{proyective}. 

A different approach to the problem of explaining the cylindicity,  an alternative to both compactified and projective approaches, comes from relaxing this condition by assuming the possibility that the physics could in principle depends on the new coordinate, making not necessarily exact the cylinder condition \cite{nocylinder}. The dependence on the fifth (or other extra dimensions) appears in regimes that have not yet been detected by experiment, such as the relevance of Minkowski's fourth dimension to mechanics was not apparent at non-relativistic speed. When dependence on an extra dimension is included then the five-dimensional Einstein equations in vacuum contain the four-dimensional ones with a general energy-momentum tensor instead of just only the electromagnetic one. 

The non-compactified approach to explain  the near-cylindricity of nature is found in the physical interpretation of the extra coordinate which like time in the Minkowski's successes of unification of Maxwell electromagnetic theory and Einstein's special relativity: it is just a geometrical generalization of the three-dimensional space by considering the time, along with space, a part of a four-dimensional space-time manifold by means of the identification $x^0=c t$ of the extra coordinate. The idea is that the extra coordinate might not be necessarily lengthlike but use transposing parameters (like the velocity of light $c$) to give it units of length. The first such proposal is the 1983 “space-time-mass” theory of Wesson \cite{wesson} who suggested that a fifth dimension $x^4$ might be associated with rest mass $m$ by means of $x^4=Gm/c^2$.  This is the mechanism used by non-compactified theories tring to justify the reason why  many experiments \cite{experiments} have been able to restrict the size of any extra dimension to below extremely small scales since all of them assume lengthlike coordinates.

The main effect of this new coordinate is that particle rest mass, usually assumed to be constant, varies with time. The variation would be small and quite consistent with experiments. This model has been studied in detail and with regard to its consequences for astrophysics and cosmology in \cite{wesson2}, \cite{wesson3} and others.

Modified theories of gravity are actually not news. What is new and different in the  non-compactified Kaluza-Klein  theories is  not so much the particular physical interpretation  attached to the new coordinate, but the fact that physics is allowed to depend on it. The first efforts in the direction of studying  the higher-dimensional Einstein equations with a general dependence on an extra coordinate and without any  preconceived notions as to its physical meaning was made by Yoshimura \cite{yoshi} (only the extra part of the space-time could depend on the extra coordinate). The general theory  has been explored by Wesson \cite{wesson2}, an others \cite{wessonetal}, and its implications for cosmology \cite{cosmology} and astrophysics \cite{astroph}.

Much effort has been made in trying to solve or at least evade the inconsistencies that emerge from general relativity as the framework for a cosmological model. The extradimensional models rise as superb possibilities on this regard. There exist many cosmological solutions for Wesson’s Space-Time-Matter (STM) theory of gravity\cite{berman}, \cite{roque}. A relation between mass variation at cosmological scales and the expansion velocity of the universe is obtained. Such a relation yields new features on Space-Time-Matter theory of gravity, which are carefully discussed in \cite{moraes}.

In cosmology, the fifth dimension can have relevant consequences, since in a homogeneous universe, the mass increases in proportion to the distance $r$ as $r^3$, and so STM differs noticeably from the
standard 4D theory of gravitation. Nevertheless, it is expected that the addition of such a fifth dimension to the usual four
does not alter noticeably the geometry of the universe at non-cosmological scales.
In fact, when we try to describe gravitational effects on non-cosmological scales, the tests of general relativity yield very small error margins in the physical measurements (precession of orbits, Iscos, radar echoes, light deflection).

At this point is where we want to show  that in the STM framework other metric solutions, rather than those from standard GR, to describe the exterior gravitational field of stellar objects are possible and compatible with the measurements. We will see that the differences arising, for example in the calculation of orbits, can be used to contrast the known 4-dimensional solutions and thus to model more accurately the physics of compact objects. For example, the differences that non-spherical metrics introduce in 4 dimensions with respect to the existence of different ISCOs \cite{lmorbits} or deformations of the celestial object with respect to the spherical symmetry can be related not so much to the deviation of the sphericity as to the existence of an extra dimension in the metric.
	
	We will see that even in the case that the dependence of metric functions on the fifth dimension is neglected new metric solutions appear from STM equations, and they can be used to describe the external gravitational field of stellar objects.


\section{The 5-dimensional metric}

Kaluza \cite{kaluza} showed that GR, when interpreted as a five-dimensional theory in vacuum (i.e. an Einstein's tensor,  with indices running over $0$ to $4$, equals to zero), contains four-dimensional general relativity in the presence of  an electromagnetic field together with Maxwell's laws of electromagnetism (the non-vacuum Einstien's equation with indices runnig from $0$ to $3$ and  an energy-momentum tensor corresponding to the electromagnetic field)

We assume that there is no five-dimensional energy-momentum tensor and hence the Einstein equations in five dimensions  are (latin indices  run over 0-4) 
\begin{equation}
\hat S_{AB}=0=\hat R_{AB} \ ,
\label{einstein5}
\end{equation}
where $\hat S_{AB}$ and $\hat R_{AB}$ denote the Einstein and Ricci tensors respectively with respect to the five-dimensional metric. The absence of matter sources is explained  in the sense  that the universe in five-dimension is assumed to be empty and the idea to explain the matter in four-dimension is as a manifestation of pure geometry in higher ones. The five-dimensional Ricci tensor and Christoffel symbols are defined in terms of the metric as in four dimensions.

Working within the Brans-Dicke \cite{bransdicke} theory\footnote{Let us note that the result obtained by Kaluza to unify the electromagnetism and gravitation arises when  an electromagnetic $A_{\alpha}$ potential is coupled to the above metric in the form $g_{\alpha \beta}+\phi^2 A_{\alpha}A_{\beta}$, and $g_{\alpha 4}=\phi^2 A_{\alpha}$.  In that case, if the scalar field $\phi$ is constant then the field equations provide directly the Einstein and Maxwell equations.}, we assume that there is no electromagnetic potential and consider the following metric (latin indices  run over 0-4, and greek ones from $0$ to $3$)
\begin{equation}
d\hat s^2=\hat g_{AB}dx^A dx^B=g_{\alpha\beta}dx^{\alpha} dx^{\beta}+ \epsilon \phi^2 d\mu^2 \ ,
\label{metrica}
\end{equation}
where the  static and axisymmetric line element of the 4-dimensional metric in a general system of  cylindric coordinates is written as follows:
\begin{equation}
ds^2=-e^{2\sigma} dt^2+e^{2 b} d\rho^2+e^{2 c} dz^2+e^{2 d}  d\varphi^2,
\label{ERglobal}
\end{equation}
and  the metric functions  in these standard coordinates depend on $\rho$ and $z$, whereas the metric function $\phi$ is allowed to depend also on $t$ and $\varphi$ coordinates. The so-called {\it cylinder condition} is assumed with respect to the independence of all the metric functions on the new coordinate $x^4\equiv \mu$. The factor $\epsilon=\pm 1$ allows us to consider a timelike or spacelike signature for the fifth dimension.

A relevant result in differential geometry known as Campbell's theorem \cite{campbell} proves that Einstein’s field equations in 4-D (with an energy-momentum tensor) can always be smoothly (if locally) embedded in the 5-D Ricci equations (vacuum field equations in five dimensions). The non-vanishing components of the five-dimensional Ricci tensor $\hat R_{AB}$ are defined in terms of the metric as in four dimensions and the calculation is standard from the metric (\ref{metrica}); the result (with the above mentioned cylinder condition) is as follows:
\begin{equation}
\hat R_{\alpha \beta}=R_{\alpha \beta}-\frac{1}{\phi}\nabla_{\beta}(\partial_{\alpha} \phi) \ , \quad \hat R_{44}=-\epsilon \phi \Box \phi \ ,
\end{equation}
$R_{\alpha \beta}$ being the four-dimensional Ricci tensor and $\Box\equiv g^{\alpha\beta}\nabla_{\beta}\partial_{\alpha}$ denotes  the standard D'Alambertian operator.
Therefore the  four-dimensional Einstein equations $S_{\alpha \beta}=8\pi G T_{\alpha \beta}$ are automatically contained in the above vacuum equations (\ref{einstein5}) if the induced {\it matter} described by $T_{\alpha \beta}$ is understood as a manifestation of pure geometry in the higher-dimensional world  (which has been called the induced matter interpretation of Kaluza-Klein theory) by the following expression:
\begin{equation}
8\pi G T_{\alpha \beta}=\frac{1}{\phi} \nabla_{\beta}(\partial_{\alpha} \phi)=R_{\alpha \beta}-\frac 12 R g_{\alpha \beta}
\label{eqcampo}
\end{equation}
and the new metric function $\phi$ satisfies a Klein-Gordon equation for a massless scalar field:
\begin{equation}
\Box \phi =0 \ ,
\label{dalambertiano}
\end{equation}
equation which is equivalent to the condition of null Ricci scalar since the trace of the energy-momentum tensor $T_{\alpha \beta}$ vanishes ($R\equiv g^{\alpha \beta}R_{\alpha \beta}=(1/\phi)\Box \phi$). 

All non-diagonal components of the Einstein tensor $S_{\alpha \beta}$ vanish except for the $S_{12}$ component, and those  energy-momentun $T_{\alpha\beta}$  tensor-components are not in general zero but depends on the behaviour of the new metric function $\phi$: if that function  depends neither on $\varphi$ nor $t$ variables then the field equations involving non-diagonal indices with values $(0)$ and/or $(3)$ (i.e. time and azimuthal angular variable) are automatically fulfilled, and hence the remaining non vanishing field equations  are exclusively those corresponding to $S_{12}$ as well as the diagonal components in the way that we shall show later.
But if we release any constraint on the function $\phi$ and allow it to depend on  $\{t,\varphi\}$ then field equations force the function not to depend simultaneously on both variables since $S_{03}=\frac{1}{\phi} \phi_{t\varphi}=0$. And in addition it must be of the form $\phi(\rho,z;t)=k_1 e^{\sigma}\hat\phi(t)$ or $\phi(\rho,z;\varphi)=k_2 e^{d}\tilde{\phi}(\varphi)$ where $\hat\phi(t)$ and $\tilde{\phi}(\varphi)$ are, in principle, arbitrary functions of the corresponding variable, and $k_1$, $k_2$ being constants.
We proceed now to show in detail such field equations for each one of  above mentioned scenarios.

\section{The new gravitational fields of compact objects}

\subsection{The field equations for the static case}

 Let us start with the first case in which the five-dimensional metric is static and axisymmetric. Hence, the new metric function $\phi$ is considered to  depends exclusively on the spatial coordinates $\rho$ and $z$, i.e.  $\phi \neq\phi(t,\varphi)$

We restrict ourselves (for simplicity, what can be implemented by means of a change of coordinates) to the case $b=c$, and without lose of generality we consider:
\begin{equation}
b=c\equiv \gamma-\sigma, \qquad  d=\ln E-\sigma \ ,
\label{nota}
\end{equation}
where $E$ and $\gamma$ are new metric functions in addition to  the remaining functions $\sigma$ and $\phi$ that must fulfill the following field equations (we have used the same combination of equations that are usually considered in the Weyl standard 4-dimension field equations, listed beside each equation): 
\begin{eqnarray}
&\Box \phi=0:& E_{\rho \rho}+E_{z z}=g E \label{ephi1}\\
& S_{11}+S_{22}:&\phi_{\rho \rho}+\phi_{z z}=-g \phi \label{ephi2}\\
& S_{00}-S_{33}:& \phi_{\rho}E_{\rho}+\phi_{z} E_{z}= g E \phi \label{ephi3} \\
& R=0:& \gamma_{\rho \rho}+\gamma_{z z}=\sigma_{\rho \rho}+\sigma_{z z}+\sigma_{\rho}\frac{E_{\rho}}{E}+\sigma_{z} \frac{E_{z}}{E}-\sigma_{\rho}^2-\sigma_{z}^2 -2 g\label{ephi4} \\
& S_{00}:&\sigma_{\rho}\left(\frac{E_{\rho}}{E}-\frac{\phi_{\rho}}{\phi} \right)+\sigma_z\left(\frac{E_{z}}{E}-\frac{\phi_{z}}{\phi}\right)=-(\sigma_{\rho \rho}+\sigma_{z z})-g \label{ephi5}
\end{eqnarray}
and the following quadrature for the metric function $\gamma$:
\begin{eqnarray}
& S_{12}:& \gamma_{z}\left(\frac{E_{\rho}}{E}-\frac{\phi_{\rho}}{\phi} \right)+\gamma_{\rho}\left(\frac{E_{z}}{E}-\frac{\phi_{z}}{\phi}\right)= 2\sigma_z\sigma_{\rho}-\frac{\phi_{z\rho}}{\phi}+\frac{E_{z\rho}}{E}-\sigma_z\frac{\phi_{\rho}}{\phi}-\sigma_{\rho}\frac{\phi_z}{\phi}\nonumber \\
& S_{11}:&  - \gamma_{z}\left(\frac{E_{z}}{E}-\frac{\phi_{z}}{\phi} \right)+\gamma_{\rho}\left(\frac{E_{\rho}}{E}-\frac{\phi_{\rho}}{\phi}\right)= \sigma_{\rho}^2-\sigma_{z}^2-\frac{\phi_{\rho\rho}}{\phi}-\frac{E_{zz}}{E}+\sigma_z\frac{\phi_{z}}{\phi}-\sigma_{\rho}\frac{\phi_{\rho}}{\phi}\nonumber \\
\label{cuadra1}
\end{eqnarray}
where $g$ is an arbitrary function which is related with the gauge of coordinates, in the sense that the election $g=0$ is equivalent to assume that the Weyl coordinates can be recovered by means of a change of coordinates, and therefore the relation for the components of the Einstein tensor is fulfilled $S_{11}+S_{22}=0$. That condition is always possible for vacuum solutions but we are now working with a non-vanishing $T_{\mu \nu}$ derived from the extra dimension.

\vskip 2mm

\subsection{Static and axisymmetric solutions}

In what follows of this work we shall consider the case $g=0$ and then the above equations deserve  some comments:

\vskip 2mm

{\bf i)} A limiting case of this set of equation is obtained with $E=\rho$ and $\phi=$cte, since  the Weyl field equations from (\ref{ephi4})-(\ref{cuadra1}) are recovered:

\begin{eqnarray}
& \gamma_{\rho \rho}+\gamma_{z z}=-\sigma_{\rho}^2-\sigma_{z}^2 \nonumber\\
&\sigma_{\rho \rho}+\sigma_{z z}+\frac{1}{\rho}\sigma_{\rho}=0 \nonumber\\
&\gamma_z=2\rho \sigma_z\sigma_{\rho}\nonumber \\
&\gamma_{\rho}=\rho (\sigma_{\rho}^2-\sigma_z^2)
\end{eqnarray}
where the integrability condition of the quadratute for $\gamma$ is just the equation for $\sigma$.

\vskip 2mm

{\bf ii)} 
A way to solve the general  set of equations ({\ref{ephi1})-(\ref{cuadra1}) may be the following: firstly we find functions $E$ and $\phi$ from equations (\ref{ephi1})-(\ref{ephi3}) which means that we are looking for two different surfaces defined by harmonic functions in "cartesian" coordinates $\rho,z$ whose gradients are orthogonal at any point. Secondly we solve equation (\ref{ephi5}) for the metric function $\sigma$, and equations (\ref{ephi4}),(\ref{cuadra1}) for $\gamma$ assuming that the integrability condition of the quadrature (\ref{cuadra1}) is fulfilled. 

If the function $\phi$ is a constant whatever $E$ be, the quadrature is fulfilled (i.e. its integrability condition) and the set of equations recovers the Weyl vacuum field equations. On the contrary, whenever the function $\phi$ is not a constant, a little bit of algebra is needed to verify that the quadrature (\ref{cuadra1}) with  the equations (\ref{ephi4})-(\ref{ephi5}) leads to the following integrability condition:
\begin{equation}
(\gamma_{\rho}-\sigma_{\rho})\left\{\frac{(\nabla \phi)^2}{\phi^2}\right\}-\frac{1}{2\phi^2}\frac{\partial}{\partial_{\rho}}\left\{ (\nabla \phi)^2\right\}=0 \ ,
\label{intcond}
\end{equation}
where $\nabla\equiv \vec e_{\rho}\partial_{\rho}+\vec e_z \partial_z $ denotes the gradient operator in a $2-$cartesian space with coordinates $\{\rho,z\}$, and so $(\nabla \phi)^2\equiv\phi_{\rho}^2+\phi_z^2$. This equation (\ref{intcond}) is fulfilled either $\phi$ is a constant or the metric function $\gamma$ verifies the following relation:

\begin{equation}
\gamma=\sigma+\frac 12 \ln\left((\nabla \phi)^2 \right)+ \xi (z) \ .
\label{gamma}
\end{equation}
where $\xi(z)$ is an arbitrary function only depending on the coordinate $z$ appearing from the integration along the coordinate $\rho$ of the equation (\ref{intcond}) since a factor $\frac{(\nabla \phi)^2}{\phi^2}\neq 0$ can be pulled out from it leading to the equation $(\gamma_{\rho}-\sigma_{\rho})-\frac{1}{2(\nabla\phi)^2}\frac{\partial}{\partial_{\rho}}\left\{ (\nabla \phi)^2\right\}=0$.

Therefore, the sets of equation to solve in the general case (with $g=0$), in addition to (\ref{gamma}), are the following:
\begin{eqnarray}
&&E_{\rho \rho}+E_{z z}=0 \quad , \quad \phi_{\rho \rho}+\phi_{z z}=0 \quad , \quad \nabla E \cdot \nabla \phi=0 \label{g01}\\
&&\sigma_{\rho \rho}+\sigma_{z z}=\nabla \sigma \cdot \frac{\nabla \phi}{\phi}-\nabla \sigma \cdot \frac{\nabla E}{E} \label{g02}\\
&&\gamma_{\rho \rho}+\gamma_{z z}=\nabla \sigma \cdot \frac{\nabla \phi}{\phi}-\left(\nabla \sigma \right)^2 \label{g03}
\end{eqnarray}
and the corresponding five-metric, for a non constant $\phi$, is
\begin{equation}
d\hat s^2=- e^{2 \sigma} dt^2+e^{2\xi}\left( \nabla \phi\right)^2\left( d\rho^2+ dz^2\right)+E^2 e^{-2\sigma}  d\varphi^2+\epsilon \phi^2 d\mu^2 \ .
\end{equation}

\vskip 2mm

{\bf iii)} The choice $E=\rho$ admits another solution for the metric function $\phi=c_1z+c_2$  which also satisfies the set of equations (\ref{ephi1})-(\ref{ephi3}) rather  than being a constant . In this particular case we end up into the following solution for the other metric functions (once  the equations (\ref{ephi4}) and (\ref{ephi5}) are solved in separated variables):
\begin{equation}
\sigma=\ln (\rho)+k\left( c_1 \frac{z^2}{2}+c_2 z \right)+c_3
\end{equation}
\begin{equation}
\gamma=\ln (\rho)+k\left( c_1 \frac{z^2}{2}+c_2 z \right)+c_3+\ln\left( \frac{c_1}{c_1z+c_2}\right)-k^2\frac{(c_1z+c_2)^4}{12 c_1^2}
\end{equation}
and therefore the corresponding metric is given by:
\begin{eqnarray}
d\hat s^2&=&- \beta_1 \rho^2 e^{k(c_1 z^2+2c_2 z)} dt^2+\beta_2 \left( \frac{c_1}{c_1z+c_2}\right)^2 e^{- k^2\frac{(c_1z+c_2)^4}{6 c_1^2}}\left( d\rho^2+ dz^2\right)+ \nonumber \\
&+&\frac{1}{\beta_1} e^{-k(c_1 z^2+2c_2 z)}  d\varphi^2+\epsilon (c_1z+c_2)^2 d\mu^2.
\end{eqnarray}
Nevertheless,  undesired behaviours of above metric advises to look for another solutions. 

Henceforth, and leaving aside the above used restrictive choice  $E=\rho$, we can hold that the more general solution of the set of equations (\ref{g02})-(\ref{g03}) is given by
\begin{equation}
\sigma=A+\ln E
\end{equation}
iff the function $A$ satisfies the equations (let us remind that also eq. (\ref{gamma}) has been taken into account):
\begin{eqnarray}
A_{\rho \rho}+A_{z z}&=&\nabla A \frac{\nabla \phi}{\phi}-\nabla A \frac{\nabla E}{E} \label{Aeq1} \\
-\xi_{zz}&=&(\nabla A)^2+\nabla A \frac{\nabla E}{E} \ .\label{Aeq2}
\end{eqnarray}
These equations can be solved, at least  for the particular case $\xi_{zz}=0$, by considering that the gradient of the function $A$ should  be a linear combination of the corresponding gradients of $E$ and $\phi$:
\begin{equation}
\nabla A =\alpha \frac{\nabla E}{E}+\beta \frac{\nabla \phi}{\phi}
\label{nablaA}
\end{equation}
for some functions $\alpha$ and $\beta$ which, after a cumbersome algebra, result to be 
\begin{eqnarray}
\beta&=&\frac{\phi^2}{2\sqrt{e^{2P}+Q^2}} \label{ab1} \\
\alpha&=&-\frac 12\pm\frac{Q}{2\sqrt{e^{2P}+Q^2}} \ ,\label{ab2}
\end{eqnarray}
where the following notation has been used
\begin{equation}
	e^{P}\equiv E\phi\frac{\phi_{z}}{E_{\rho}} \quad , \qquad Q\equiv k(z)\pm\int\frac{\partial}{\partial z}\left( e^P \right) d \rho
	\end{equation}
	with $k$ being an arbitrary function of $z$, and the function $A$ being integrable from equation (\ref{nablaA}).

	 An explicit solution for the five-dimensional metric can be obtained by assuming now a particular choice of the functions $E$ and $\phi$: that is the 
	 	 remaning discussion  related with the existence of solution for the pair of metric functions $E$ and $\phi$ with a reasonable behaviour (equations (\ref{g01})). We are looking for a couple of  solutions of the equation $v_{\rho \rho }+v_{zz}=0$ whose gradients are orthogonal, and those functions vanish asymptoticaly at infinity. As is well known, $v$ being  an harmonic  solution then a solution $w$ of the Cauchy-Riemann conditions: $w_z=v_{\rho}$, $w_{\rho}=-v_z$  is an harmonic solution as well. In addition, those conditions lead to  both function $v$ and $w$ to hold that $v_{\rho}w_{\rho}+v_z w_z=0$.
	 	\footnote{The Cauchy-Newman problem:
	 	$
	 	v_{\rho \rho }+v_{zz}=0 \ ,\ v(\rho, z)\left|_{\rho \rightarrow 0}\right.=0
	 	$
	 	has the following general solution (with a vanishing behaviour at infinity)
	 	\begin{equation}
	 	v=\rho\sum_{n=0}^{\infty} \frac{h_n}{r^{n+2}} C_n^{(1)}(\omega) \ , \quad \omega\equiv\cos\theta\equiv \frac{z}{r}=\frac{z}{\sqrt{\rho^2+z^2}} \nonumber
	 	\end{equation}
	 	with arbitrary constants $h_n$ and $C_n^{(1)}(\omega)$ being Gegenbauer polynomials of degree $n$. And  $w$ is the solution of the Cauchy-Riemann conditions from the harmonic function $v$:
	 	\begin{equation}
	 	w=\sum_{n=0}^{\infty} \frac{h_n}{r^{n+1}} \left[ C_{n-1}^{(1)}(\omega)-\omega C_n^{(1)}(\omega) \right]\nonumber
	 	\label{uwedoble1}
	 	\end{equation}
	 	Therefore the election of our metric functions $E$ and $ \phi$ 
	 	could be
	 	$
	 	E=c_4\rho+v\ , \ \phi=c_3+c_4z+w
	 	$ for any $c_3$, $c_4$ arbitrary constants.}
	 	
	For such a choice of the functions $E$ and $\phi$, i.e, assuming for both functions a Cauchy-Riemman condition, the we can manage an explicit expression for the metric functions since $E_{\rho}=\phi_z$ and then
\begin{eqnarray}
\beta&=&\frac{\phi^2}{2\sqrt{E^2\phi^2+\left(k\pm \frac{E^2-\phi^2}{2}\right)^2}} \label{abej1} \\
\alpha&=&-\frac 12\pm\frac{k\pm \frac{E^2-\phi^2}{2}}{2\sqrt{E^2\phi^2+\left(k\pm \frac{E^2-\phi^2}{2}\right)^2}} \ ,\label{abej2}
\end{eqnarray}
with $k$ being an arbitrary constant, and the function $A$ can be obtained by a simple integration:
\begin{equation}
A =\int \left(\alpha \frac{E_z}{E}+\beta \frac{\phi_z}{\phi}\right) \ dz= \int \left(\alpha \frac{E_{\rho}}{E}+\beta \frac{\phi_{\rho}}{\phi}\right) \ d\rho \ ,
\end{equation}
up to an arbitrary constant $C$:
\begin{equation}
A=C-\ln E+\frac 12 \ln \left( E^2+\phi^2\mp 2k+\sqrt{(E^2+\phi^2\mp2k)^2\pm8kE^2}\right)\equiv C-\ln E+\frac 12 \ln H ,
\label{A}
\end{equation}
leading to the following metric components:
\begin{eqnarray}
g_{00}&=&-E^2 e^{2A}=-e^{2C} H \label{gij1} \\
g_{11}=g_{22}&=&e^{c_1z+c_2}(\nabla \phi)^2\label{gij2}\\
g_{33}&=&e^{-2A}=e^{-2C}\frac{E^2}{H} \label{gij3}\\
g_{44}&=&\epsilon \phi^2 \label{gij4}
\end{eqnarray}

\subsection{The spherically symmetric solutions}

The spherial symmetric case is not included in the family of  solutions above obtained. Therefore we proceed to calculate those solutions in another system of coordinates $\{r,\theta \}$ rather than those previously used. The more general line element with this symmetry in five dimensions is written as follows:

\begin{equation}
ds^2= -A^2 dt^2+B^2 dr^2+r^2 (d\theta^2+\sin \theta^2 d\varphi^2)+\epsilon \phi^2 d\mu^2 \ ,
\end{equation}
where the metric functions $A=A(r)$, $B=B(r)$ and $\phi=\phi(r)$ only depend on the radial coordinate.

The field equations, with the notation $a\equiv \ln A$ , $b\equiv \ln B$ are the following ones:
\begin{eqnarray} -\frac{B^2}{r^2}-2\frac{b^{\prime}}{r}+\frac{1}{r^2}&=&-a^{\prime}\frac{\phi^{\prime}}{\phi} \label{sph1}\\
\frac{B^2}{r^2}-2\frac{a^{\prime}}{r}-\frac{1}{r^2}&=&\frac{\phi^{\prime \prime}}{\phi}-b^{\prime}\frac{\phi^{\prime}}{\phi} \label{sph2}\\
-a^{\prime \prime}-a^{\prime 2}+b^{\prime} a^{\prime}  +\frac{b^{\prime}-a^{\prime}}{r}&=&\frac{\phi^{\prime}}{r \phi}    \label{sph3}  
\end{eqnarray}
in addition to the equation (\ref{dalambertiano}) for the new metric function $\phi$ which leads to
\begin{equation}
\phi^{\prime}=\frac{K}{r^2} e^{b-a} \label{phiprima} \ .
\end{equation}
 The sum of equations (\ref{sph1}) and (\ref{sph2}) leads to
\begin{equation}
\phi^{\prime \prime}-\phi^{\prime}(a^{\prime}+b^{\prime})=-\frac{2}{r}(a^{\prime}+b^{\prime}) \phi \label{phi2prima} \ ,
\end{equation}
and then, in combination with (\ref{phiprima}) we get 
\begin{equation}
\phi=\frac{(r a^{\prime}+1)K e^{b-a}}{r^2(a^{\prime}+b^{\prime})} \label{phisph} \ .
\end{equation}

Let us note here that the Schwarzschild solution of the four-dimensional case can be recovered if we consider $\phi=cte$, since equation (\ref{phi2prima}) would imply $a^{\prime}=-b^{\prime}$ and then equations (\ref{sph1}) and (\ref{sph2}) are equals and lead, within  the equation (\ref{sph3}), to obtain the known metric functions $a$, $b$ of the spherical symmetric solution in the standard spherical coordinate of Schwarzschild: $a=\frac 12 \ln(1-2m/r)$.

Going back to the previous equations for five-dimensional metric all the equations (\ref{sph1})-(\ref{sph3}) are identicals:
\begin{equation}
a^{\prime 2}-a^{\prime}\left(b^{\prime}-\frac 1r +\frac{B^2}{r} \right)-\left(2\frac{b^{\prime}}{r}-\frac {1}{r^2} +\frac{B^2}{r^2} \right)=0 \label{cuadratica}
\end{equation}
if we make  use of the null Ricci scalar, i.e.:
\begin{equation}
a^{\prime \prime}+a^{\prime 2}-\frac{B^2}{r^2}+b^{\prime} a^{\prime}  -2\frac{b^{\prime}-a^{\prime}}{r}+\frac{1}{r^2}=0 \ . \label{ricci}
\end{equation}
Therefore the remaining field equations  to obtain the spherical solution are (\ref{ricci}) and (\ref{cuadratica}). The combination of both equations leads to
\begin{equation}
B^2=-1-r\frac{a^{\prime \prime}}{a^{\prime}} \label{B2}
\end{equation}
and equation (\ref{ricci}) turns out to be, with the notation $y\equiv r a^{\prime}$
\begin{equation}
y^{\prime 2}(4+3y)-y y^{\prime \prime}(2+y)+y^2 \frac{y^{\prime}}{r}(2y+1)=0 \ ,\label{ecuay}
\end{equation}
whose solution is the following
\begin{equation}
q=\frac{r^2 y^2}{cy^2+2y(c+1)+1}\left(\frac{y\sqrt{c(c+1)+1}-1-y(c+1)}{y\sqrt{c(c+1)+1}+1+y(c+1)} \right)^{\frac{c+1}{\sqrt{c(c+1)+1}}}\label{tocho}
\end{equation}
for any constants $q>0$  and $c$.

The case $c=-1$ allows to clear the function $y$ (${\displaystyle y\equiv a^{\prime} r=\sqrt{\frac{q}{q+r^2}}}$), and thus we are able to calculate available metric functions $A^2\equiv e^{2a}$, $B^2\equiv e^{2b}$ (from equation (\ref{B2})) and $\phi$ (from equation \ref{phisph})) as follows:
\begin{eqnarray}
A^2&=& \frac{r^2}{4q(\sqrt q+\sqrt{r^2+q})^2} \\
B^2&=&\frac{r^2}{r^2+q} \\
\phi&=&\frac{2K}{r}\left(\sqrt q+\sqrt{r^2+q}\right)
\label{spheric5D}
\end{eqnarray}

\subsection{Time-dependent solutions}

We are going finally to show  that non-static five-dimensional metrics with time dependence are compatible with the set of equations. Let us consider the new metric function $\phi$   depending on the variable $t$, in addition to the spatial coordinates,  i.e.  $\phi =\phi(\rho,z;t)$. As we already commented above the field equations do not allow  a dependence of that metric function  on both coordinates $\varphi$ and $t$  simultaneously, and in addition it has to be of the form $\phi(\rho,z;t)=k_1 e^{\sigma}\hat \phi(t)$ with  $k_1$  being constant.

We hold the line element (\ref{ERglobal}) and restrict ourselves, as well as in the previous case,  to the case $b=c$, introducing the metric functions $E$ and $\gamma$ defined in (\ref{nota}). Since the equation (\ref{dalambertiano}) leads to $\hat \phi_{tt}/\hat \phi=\alpha$ then the metric function $\phi$   must be assumed to be: 
\begin{equation}
\phi=k_1e^{\sigma}\left( c_1 e^{\sqrt \alpha t}+c_2 e^{-\sqrt \alpha t}\right)
\end{equation}
where $c1,c2,k$ and  $\alpha$ being arbitrary  constants, and the metric functions $\sigma$, $E$, $\gamma$ are considered to depend exclusively on the spatial coordinates $\{\rho, z\}$ . Therefore only those three metric functions remaing unsolved, $E$, $\sigma$ and $\gamma$ from the following field equations: 
\begin{eqnarray}
&E_{\rho \rho}+E_{z z}=0 \label{settime1}\\
& \sigma_{\rho \rho}+\sigma_{z z}=\gamma_{\rho \rho}+\gamma_{z z}=-\sigma_{\rho}^2-\sigma_{z}^2 \label{settime2} \\
& \sigma_{\rho}\frac{E_{\rho}}{E}+\sigma_{z} \frac{E_{z}}{E}=\sigma_{\rho}^2+\sigma_{z}^2=\alpha e^{2\gamma-4\sigma} 
\label{settime3}
\end{eqnarray}
as well as the following quadrature for the metric function $\gamma$:
\begin{eqnarray}
& \gamma_{z}\left(\frac{E_{\rho}}{E}-\sigma_{\rho} \right)+\gamma_{\rho}\left(\frac{E_{z}}{E}-\sigma_{z}\right)= -\sigma_z\sigma_{\rho}+\frac{E_{z\rho}}{E}-\sigma_{z\rho}\nonumber \\
&  - \gamma_{z}\left(\frac{E_{z}}{E}-\sigma_{z} \right)+\gamma_{\rho}\left(\frac{E_{\rho}}{E}-\sigma_{\rho}\right)= -\sigma_{\rho}^2-\sigma_{\rho\rho}-\frac{E_{zz}}{E}
\label{cuadra2}
\end{eqnarray}

Firstly, we can  consider a restriction,  if we take for the metric function $E=\rho$ as a solution of (\ref{settime1}), and then the set of remaining equations turns out to be
\begin{eqnarray}
&\triangle \sigma=0 \label{time1}\\
& \sigma_{\rho \rho}+\sigma_{z z}=\gamma_{\rho \rho}+\gamma_{z z}=-\sigma_{\rho}^2-\sigma_{z}^2 \label{time2}\\
& \sigma_{\rho}=\rho\left(\sigma_{\rho}^2+\sigma_{z}^2\right)=\rho\alpha e^{2\gamma-4\sigma} \label{time3} \\
& \gamma_{z}\left(\frac{1}{\rho}-\sigma_{\rho} \right)-\gamma_{\rho}\sigma_{z}= -\sigma_z\sigma_{\rho}-\sigma_{z\rho}\nonumber \\
&  \gamma_{z}\sigma_{z}+\gamma_{\rho}\left(\frac{1}{\rho}-\sigma_{\rho}\right)= -\sigma_{\rho}^2-\sigma_{\rho\rho}
\end{eqnarray}
and a simple solution is obtained if we take $\sigma_{\rho}=1/\rho$, $\sigma_z=0$,  since the quadrature for $\gamma$ disappear, equation (\ref{time1}) is automatically fullfiled, and equations (\ref{time2})-(\ref{time3}) requiere
\begin{equation}
\gamma_{\rho\rho}+\gamma_{zz}+\frac{1}{\rho^2}=0 \ , \qquad 1=\alpha  \rho^2 e^{2\gamma-4\sigma}
\end{equation}
whose general solution is given by the following metric functions:
\begin{equation}
\sigma=C+\ln(\rho) \ , \qquad \gamma=2C+ \ln\left(\frac{\rho}{\sqrt{\alpha}}\right)
\end{equation}
and hence the five-dimensional metric looks like (for any constants $C$, $k$, $c_1$, $c_2$ and $\alpha\neq0$)
\begin{equation}
d\hat s^2=-e^{2C}\rho^2 dt^2+\frac{e^{2 C}}{\alpha}\left( d\rho^2+ dz^2\right)+e^{-2 C}  d\varphi^2+\epsilon k^2e^{2C} \rho^2\left( c_1 e^{\sqrt \alpha t}+c_2 e^{-\sqrt \alpha t}\right)^2 d\mu^2
\end{equation}

Nevertheless, this is not the general solution of our set of equations (\ref {settime1})-(\ref{settime3}) and (\ref{cuadra2}). From the second identity of (\ref{settime3}) we have that 
\begin{equation}
\gamma=2 \sigma+\frac 12 \ln\left(\frac{(\nabla \sigma)^2}{\alpha}\right) 
\label{gammat}
\end{equation}
and the integrability conditions of (\ref{cuadra2}) is fulfilled iff $\sigma_{\rho \rho}+\sigma_{zz}=-(\nabla \sigma)^2$. Hence the remaining equations are
\begin{eqnarray}
&E_{\rho \rho}+E_{z z}=0 \label{rem1}\\
& \sigma_{\rho \rho}+\sigma_{z z}=-\sigma_{\rho}^2-\sigma_{z}^2 \label{rem2} \\
& \sigma_{\rho}\frac{E_{\rho}}{E}+\sigma_{z} \frac{E_{z}}{E}=\sigma_{\rho}^2+\sigma_{z}^2
\label{rem3}\\
&\gamma_{\rho \rho}+\gamma_{z z}=-\sigma_{\rho}^2-\sigma_{z}^2 \label{rem4}
\end{eqnarray}
Now, the unique solution of the couple of equations (\ref{rem2}) and (\ref{rem3}) is $\sigma=\ln E$ and then equations (\ref{rem1}) and (\ref{rem2}) are equivalent and (\ref{rem3}) is an identity. In addition  the equation (\ref{rem4}) with (\ref{gammat})
leads to the  equation:
\begin{equation}
 (\partial_{\rho \rho}+\partial_{z z})\left[\ln\left((\nabla \sigma)^2 \right)\right]=2(\nabla \sigma)^2 \ ,
\label{remf2}
\end{equation}
which once again  is an identity since, as  it is known, any solution of (\ref{rem1}) fulfills
\begin{equation}
(\partial_{\rho \rho}+\partial_{z z})\left[\ln\left((\nabla E)^2 \right)\right]=0
\end{equation}
and therefore the general solution for a five-dimensional metric  in this case is given by
\begin{equation}
d\hat s^2=-E^2 dt^2+\frac{(\nabla E)^2}{\alpha}\left( d\rho^2+ dz^2\right)+
 d\varphi^2+\epsilon k^2 E^2\left( c_1 e^{\sqrt \alpha t}+c_2 e^{-\sqrt \alpha t}\right)^2 d\mu^2,
 \label{metricatemporal}
\end{equation}
for arbitray constants $c_1$, $c_2$, $k$ and $\alpha\neq 0$, where $E$ is a solution of the equation (\ref{rem1}).

\section{Old and/or new physics}

\subsection{The geodesics equations}

We obtain the equations of motion by minimizing the five-dimensional interval $d\hat s^2$ (\ref{metrica}) leading to a version of the geodesic equation \cite{geod5}:
\begin{equation}
\frac{d^2x^A}{d\hat s^2}+\hat\Gamma^A_{BC}\frac{d x^B}{d\hat s}\frac{d x^C}{d\hat s}=0
\label{geod5D}
\end{equation}
where the five dimensional Christoffel symbols $\Gamma^A_{BC}$ are defined identically to the four dimensional case but with respect to the extended metric. The component $A=4$ of the geodesic equation takes the form \cite{geod5} 
\begin{equation}
\frac{dB}{d\hat s}=\frac 12\frac{\partial \hat g_{CD}}{d\mu}\frac{d x^C}{d\hat s}\frac{d x^D}{d\hat s}
\end{equation}
where $B$ is a scalar function ${\displaystyle B\equiv \phi^2\left( \frac{d \mu}{d \hat s}\right)}$ that is a constant of  motion in the particular case that we are studing since the metric $\hat g_{AB}$ does not depend on the extra coordinate $x^4\equiv \mu$ (since $dB/d\hat s=0$). The definition of $B$ together with the form of the metric (\ref{metrica}) allow us to relate the five-dimensional interval with the four-dimensional one by means of ${\displaystyle d\hat s=\frac{1}{\sqrt{1-\epsilon B^2/\phi^2}} ds}$, and then the equations of geodesics take the following form \cite{geod5}
\begin{equation}
\frac{d^2 x^\nu}{ds^2}+\Gamma^{\nu}_{\alpha \beta} \frac{dx^{\alpha}}{ds}\frac{dx^{\beta}}{ds}=\frac{\kappa}{1-\kappa}\left[ \frac{1}{\phi}\frac{\partial \phi}{\partial x^{\nu}}-\frac{1}{\phi}\frac{d \phi}{ds} \frac{dx^{\nu}}{ds}\right] \ ,
\label{geodesics5D}
\end{equation}
where the notation ${\displaystyle \kappa\equiv \frac{\epsilon B^2}{\phi^2}}$ has been  used.

Let us consider now those geodesics located on the equatorial plane, i.e., $z=$cte.
The stationary and the axial symmetry allow to integrate the geodesic equations (\ref{geodesics5D}) for the variables $t\equiv x^0$ and $\varphi\equiv x^3$ respectively to get:
\begin{eqnarray}
\frac{dt}{ds}&=&\frac{\hat h}{g_{00}\sqrt{1-\kappa}}\\
\frac{d\varphi}{ds}&=&\frac{\hat l}{g_{33}\sqrt{1-\kappa}} \label{varphi}
\end{eqnarray}
where $\hat h$, $\hat l$ are both constants of motion associated to the Killing vectors of the isometries.

The remaining geodesic on the equatorial plane is
\begin{equation}
\frac{d^2\rho}{d s^2}
+\left(\frac{d \rho}{d s}\right)\frac{d}{d s}\left[ \ln (\sqrt{1-\kappa})\right]+\frac 12\left(\frac{d \rho}{d s}\right)^2 \partial_{\rho}\ln(g_{11})=C
\end{equation} 
where ${\displaystyle C\equiv \frac{\kappa}{1-\kappa} \frac{\phi_{\rho}}{\phi}+\frac{1}{2g_{11}(1-\kappa)} \left[ \frac{\partial_{\rho}g_{00}}{g_{00}^2}+\frac{\partial_{\rho}g_{33}}{g_{33}^2} \right] }$

This equation can be solved for ${\displaystyle \left(\frac{d \rho}{d s}\right)}$ to obtain:
\begin{equation}
\left(\frac{d \rho}{ds}\right)^2=\frac{1}{1-\kappa}\left[ -\kappa+\delta_{\pm}-\frac{\hat h^2}{g_{00}}-\frac{\hat l^2}{g_{33}}\right]\frac{1}{g_{11}}\equiv \frac{V(\rho)}{1-\kappa}
\label{rhogeod}
\end{equation}

Let us note that this result, for the case $\phi=$ cte, is already obtained in \cite{kepler}. In addition,  this expression for the remaining geodesic can be obtained from the conservation of the norm of the five-dimensional tangent vector ${\displaystyle \hat u^A\equiv \frac{dx^A}{d\hat s}}$,  $\hat u^A \hat u_A=\delta_{\pm}$ leading to the following expression for the norm of the four-dimensional tangent vector $u^{\nu}$ (the indices $\alpha$, $\nu$  are in notation of sum running from $0$ to $3$):
\begin{equation}
||u||\equiv u^{\nu} u_{\nu}\equiv g_{\alpha\alpha}\left(\frac{dx^{\alpha}}{d s} \right)^2=\frac{\delta_{\pm}-\kappa}{1-\kappa}
\label{modulos}
\end{equation}
Let us observe at this point that the type of the tangent vectors $u$ and $\hat u$ are equals (for any sign of $\epsilon$) since $1-\kappa>0$ as well as $|\phi^2|>|B^2|$, ($\kappa<1$) because ${\displaystyle \left(\frac{ds}{d\hat s}\right)^2=1-\kappa}$, and consequently we have that $\delta_{\pm}>0 (\delta_{\pm}<0) \rightarrow ||u||>0 (||u||<0)$.

The geodesic (\ref{rhogeod}) corresponds to the orbit of a particle moving on the equatorial plane by means of (\ref{varphi}):
\begin{equation}
\left(\frac{du}{d\varphi}\right)^2=V(1/u)\frac{u^4 g_{33}^2}{\hat l^2} \ , \label{orbit}
\end{equation}
where the notation $u\equiv1/\rho$ has been used.

The integration of the geodesic (\ref{orbit}), or equivalently the relativistic Binet equation:
\begin{equation}
\frac{d^2u}{d\varphi^2}=\frac 12 \frac{d}{du}\left( V(1/u)\frac{u^4 g_{33}^2}{\hat l^2}\right)\label{binet}
\end{equation}
 leads to an elliptic integral, which can be obtained from this autonomous equation (\ref{binet}) as well as from (\ref{orbit}), and provides us with the orbit of the test particle:
 \begin{equation}
 \varphi+c_2=\int\frac{du}{\sqrt{c_1+u^4Vg_{33}^3/\hat l^2}}
 \end{equation} 
 
The explicit resolution of the above integral is rather complicated and hence, we will instead make use of a procedure to obtain information of the orbit. It consists on a comparison with the one-dimensional equivalent problem describing the  movement of a test particle of mass $m$ into a classical effective gravitational potential(remember that the geodesic is considered to be in a $\theta$-constant plane): 
\begin{equation}
\Phi_{eff}=\frac{J^2}{2mr^2}-GM\frac{m}{r}+V_{p}
\end{equation}
for a perturbed  (generalized) Kepler problem \cite{goldstein}, with $V_p$   a  perturbative potential. We shall develop in power series of $u$ the orbit (\ref{orbit}) and relate the arising coefficients with those of the classical perturbed orbit (with energy $\bar E$, and angular momentum $J$):
\begin{equation}
\left(\frac{du}{d\varphi}\right)^2=-u^2-\frac{2m}{J^2}(\bar  E+GMmu-V_p)\label{orbitclass}
\end{equation}
This is the mechanism already used to identify the Schwarzschild spherical solution in $4$-dimensional space-time from a perturbed Kepler problem with a potential $V_p=-\alpha/r^3$.
In \cite{kepler} that perturbed potential corresponding to  the relativistic LM solutions \cite{LM} is explicitly calculated. In \cite{kepler} this perturbed potential is used to calculate corrections to the quadrupole moment of Sun by relating with the perihelion advance of Mercury. In \cite{lmorbits} the corrections to the ISCOs associated to the different multipole moments are calcuated in comparisson with those of the Schwarzschild solution.

Therefore, if we expand in powers series of $u$ (assuming as it is so that  the medium lengths or the orbits allows to consider $u<<1$) the expression (\ref{orbit}) corresponding to the spherical 5-D solution (\ref{spheric5D}), then we can interpret the relativistic corrections provided by this vacuum solution to the orbital motion as the perturbation of a classical Newtonian potential of the type 
\begin{equation}
V_p^{sph}=\frac{1}{r^3}\left( \frac{\sqrt{q^3}}{\hat l^2} (24 \hat h^2 q+\delta_{\pm})\right)+\frac{1}{r^4}\left(- q (\hat 1+\frac{2\delta_{\pm} q}{3 \hat l^2})\right)+O(u^5) \ .
\label{Vpsph}
\end{equation}
where the  new constant of motion introduced from the five dimension $B$ has been taken such that  ${\displaystyle B^2=\frac{4 K^2}{3 \epsilon}(12 \hat h^2 q+\delta_{\pm})}$, and $\hat l$, $\hat h$ are the constants of motion (\ref{varphi}) associated to this static and axisymmetric five-dimensional metric.

The expression (\ref{Vpsph}) means that the gravitational effects of the spherically symmetric STM solution obtained above can be described, in the same way used with the Schwarzschild solution, by means of a classical perturbation of the Newtonian potential. Whereas that perturbative potential for Schwarzschild is proportional to $1/r^3$, the STM solution provides additional corrections to that perturbative potential of that same order.

Alternatively we can show that the orbit on the equatorial plane
corresponding to this spherical STM solution  is given by the expression
\begin{equation}
\left(\frac{du}{d\varphi}\right)^2=\frac{2\delta_{\pm}}{3 \hat l^2}+\frac 23\frac{\sqrt q (24 \hat h^2 q+\delta_{\pm})}{\hat l^2} u -u^2+\frac{\sqrt{q^3}}{\hat l^2} (24 \hat h^2 q+\delta_{\pm}) u^3+O(u^4)  \ ,
\label{orbitsph5D}
\end{equation}
and, in order to recover  the same equation for the orbit of Schwarzschild , i.e.;
\begin{equation}
\left(\frac{du}{d\varphi}\right)^2=\frac{ \delta+ h^2}{ l^2}-\frac{2M \delta}{ l^2} u -u^2+2M u^3\label{orbitschwarz}
\end{equation}
then the parameters  $\hat l$, $\hat h$ must be choosen as  some combination of those corresponding  to the $4$-dimensional spherical metric, $h^2\equiv -\delta(\frac{2E}{mG}+1)$, $l^2\equiv-\frac{\delta J^2}{m^2 G}$, which are related with angular momentum $J$ and energy $E$ of the orbital motion of a test particle of mass $m$:
\begin{equation}
\hat l^2=\frac{2 \delta_{\pm}}{3}\frac{l^2}{\delta+h^2} \ , \quad \hat h^2= \frac{\delta_{\pm}}{24 q} \left(-1- \frac{2M\delta}{\sqrt{q}(\delta+h^2)}\right)
\end{equation}
This choice of parameters recovers the Schwarzschild orbit (\ref{orbitschwarz})  up to order $O(u^2)$, just is to say the newtonian terms, leading to the contribution of order $u^3$: ${\displaystyle -\frac{3M\delta q}{l^2} u^3 }$. In addition, if  we want to recover the same expression at order $u^3$  the parameter $q$ of the metric in five dimensions must be taken  ${\displaystyle q=-\frac 23 \frac{l^2}{\delta}}$.

Consequently, the $5$-dimensional metric  gives us  possible differences with respect to  the relativistic Schwarzschild correction to the classical orbital motion, both if the parameter $q$ of the metric remains free or by taking int account the successive corrections in higher orders of $u$. These corrections are comparable with those that appear in $4-$dimensional GR when we work with non-spherical metrics \cite{lmorbits}.

\subsection{Variation of rest mass} 

The definition of geodesic in a five-dimensional metric (\ref{geod5D}) leads to an equation formaly in the same way as in the four-dimensional theory. Nevertheless, when the equation of geodesic is developed in terms of the four-dimensional interval $d\hat s$, the resulting equations for the spatial components (\ref{geodesics5D}), show a right-hand side that in general contains non-vanishing terms involving the spatial velocities and the extra part of the metric. These terms, bringing to the surface the existence of a fifth force (from the viewpoint of the four-dimensional standard GR) provide an explanation of why galaxies with large velocities could not necessarily travel along the known four-dimensional geodesics.

 An exciting prediction of five-dimensional general relativity arises if we interpret the fifth dimension physically in the context of STM theory, where the extra dimension $\mu$ is related to the particle rest mass. 
 As already mentioned, the embedding of spacetime in a Ricci-flat 5D
 manifold guarantees that the line element $d\hat s$ will contain $ds$. Accordingly, particles moving on null paths in 5D ($d\hat s^2=0$) will appear as massive
 particles moving on timelike paths in 4D (or spacelike one depending on the sign of $\epsilon$). Consequently, massive particles appear in 4D with $m = m(s)$. This
 suggests an intriguing possibility, that the presence of the fifth dimension could in principle be detected as a variation of the (rest) mass of a particle with proper time.

 In such a case that variation in rest mass with time can be calculated as follows. If we define a five-velocity ${\displaystyle \hat v^A\equiv \frac{dx^A}{d \hat s}}$, which is related to the usual four-velocity $v^{\alpha}$ by ${\displaystyle v^{\alpha}=\hat v^{\alpha}(\frac{1}{\sqrt{1-\kappa}})}$ then comoving objects (in comoving spatial coordinates such that $\hat v^i=0$) the remaining components of the geodesic equation are
 \begin{eqnarray}
&& \frac{d \hat v_0}{d \hat s}+\frac{\hat v_0 \hat v_4}{g_{00}}\partial_4 g_{00}-\frac 12 \frac{\hat v_4^2}{g_{00}}\partial_0 g_{44}=0\nonumber \\
&& \frac{d \hat v_4}{d \hat s}+\frac{\hat v_0 \hat v_4}{g_{44}}\partial_0 g_{44}-\frac 12 \frac{\hat v_0^2}{g_{44}}\partial_4 g_{00}=0
 \label{mdet}
 \end{eqnarray}
 If we reduce ourselves to  a five-dimensional metric whose components do not depend on the extra coordinate (the cylinder condition) but on time like the one obtained in the previous section (\ref{metricatemporal}) we can integrate those geodesics  taking into account the relation (for the time-like case $\delta_{\pm}=-1$):
 \begin{equation}
 g_{00}\hat v_0^2+g_{44}\hat v_4^2=\delta_{\pm}=-1
 \end{equation}
Hence, the non-vanishing components of the five-velocity are
\begin{equation}
\hat v_0=\frac{\sqrt{-g_{44}-\nu^2}}{\sqrt{g_{00}g_{44}}} \ , \quad \hat v_4=\frac{\nu}{g_{44}}
\end{equation}
for any  non-zero arbitrary  constant $\nu$. Obviously the ratio of these components gives us the rate of change with time of the extra coordinate, since ${\displaystyle \hat v_0\equiv \frac{dx^0}{d \hat s}=\frac{c d t}{d \hat s}}$ and ${\displaystyle \hat v_4\equiv \frac{dx^4}{d \hat s}=\frac{d \mu}{d \hat s}}$:
\begin{equation}
\frac{d\mu}{dt}=\frac{\hat v_4}{\hat v_0}=\frac{\nu \sqrt{-g_{00}}}{\sqrt{g_{44}}\sqrt{\nu^2+g_{44}}}
\end{equation}
whose integration yields to the time-dependent expression for the rest mass:
\begin{equation}
\mu(t)=\mu_0\ln\left[\nu^2\left(\frac{c_1e^{2\sqrt{\alpha} t}-c_2}{c_1e^{2\sqrt{\alpha} t}+c_2}-\frac{1}{\mu_0}\sqrt{\frac{E^2}{\alpha \nu^2}+\frac{e^{2\sqrt{\alpha} t}}{\alpha \epsilon k^2(c_1e^{2\sqrt{\alpha} t}+c_2)^2}} \right) \right]
\label{emedet}
\end{equation}
where ${\displaystyle \mu_0\equiv -\frac{c \nu}{2k \sqrt{\alpha \epsilon}\sqrt{-\nu^2c_1c_2}}}$.

In order to evaluate whether the above variation of the rest mass  is an event available for any time, it is  convenient to rewrite (\ref{emedet}) as follows:
\begin{equation}
\mu(t)=\mu_0 \ln \left[ \nu^2 \left( \frac{T-c_2}{T+c_2}- \sqrt{n_1+ \frac{4 T n_2  c_2}{(T+c_2)^2} } \right) \right]
\label{emedetrev}
\end{equation}
where ${\displaystyle T\equiv c_1 e^{2\sqrt{\alpha} t}}$ is a reparametrized time, $c_1\equiv T_0$ being a parameter with the meaning of initial time, and the notation ${\displaystyle n_1\equiv \frac{E^2}{\alpha \mu_0^2\nu^2}} $, ${\displaystyle n_2\equiv \frac{1}{4 \alpha \mu_0^2 c_1 c_ 2 \epsilon k^2}} $ has been used. The domain of the function $\mu(t)$ is resctricted by the following conditions:
\begin{equation}
(T-c_2)^2 > n_1(T+c_2)^2+4 n_2 c_2 T > 0
\label{desigualds}
\end{equation}
The curves in $T$ from above inequalities are polynomials of second degree, and the fulfillment of both conditions requieres simultaneously the  following restrictions on the parameters: 
\begin{equation}
0<n_1<1 \ , \quad -1< n_2 < -n_1 \ , \quad -1 \leq\frac{n_2}{n_1} \leq 0
\end{equation}
These conditions are impossible to be verified, and therefore given this incompatibility we conclude that the existence of a variation of the rest mass only happens in a limited time interval.

\section{Conclusions}

Just as the existence of $c$, a fundamental constant of GR, suggests a coordinate to be defined as $ct$, the existence of a second fundamental constant on GR, namely $G$, might suggest
$Gm/c^2$ to be defined as a coordinate. In that five-dimensional manifold the STM theory is developed and the vacuum five-dimensional Einstein equations are showed to include the standard four-dimensional field  equations endowed with an energy-momentum tensor which depends on the new metric function added.
It is  important to emphasize that the fifth dimension should not be understood as a new extra-spatial dimension but as an artifice to measure the physical gravitational phenomena by carrying out an extension of the metric coordinates in the same way that Minkowski and Einstein incorporated time into the metric of space, giving rise to the theory of special relativity.

The exterior of a self-graviting compact object is no longer vacuum but {\it contains "matter"} represented by that energy-mmentum tensor arising from the extra dimension included in the metric. This is of course no real matter but it acts like it would really exists in the sense that this induced {\it  matter} is contributing to perturb the space-time and the gravitational field of the compact object has to account for it through the new field equations arising.
 
In addition to  the external gravitational field we could also approach the study of the solution inside the compact object by solving the equations with an energy-momentum tensor, describing the desired source, attached to the one induced by the five-dimensional geometry. Although that was not the purpose of this work (see other works like \cite{strangestars}).

We have been  able to obtain static and axially symmetric solutions in a five-dimensional STM theory suitable  to describe the gravitational  field of different astrophysical scenarios. 

We have also obtained a spherically symmetric family of solutions. As is known  Birkoff's theorem does not hold in five dimensions,  and another solutions have been already obtained. The calculation of the geodesic (for the spherical symmetric case) in a constant plane allow us to distinguish diferences with respect to that obtained in four-dimensional standard gravity. As well, those different geodesics that are obtained in four-dimensional gravity from metrics that are no spherically symmetric can be compared with the five-dimesional case within  their effective potentials. Hence, we can conclude that the deviations in  four-dimensional gravity of non-spherically symmetric compact objects with respect to the spherical case is compatible with an explanation related with an extra dimension rather than the symmetry of the metric itself.

The extension to time-dependent case has already been addresed and we heve obtained metrics with time dependency in the new metric function associated to the added dimension. The variation of rest mass in this case is studied. In \cite{LWP} Lui, Wesson and Ponce obtained one class of these solutions but within the sperically symmetric case. In \cite{portilla} a revision of the concept of mass of a particle in GR is developed, and it is shown that  dark matter and dark energy can be explained as a gravitational effect. 

The gravity exerted by dark matter is the main explanation for why galaxies rotate so fast, why galaxies orbit clusters so fast, why gravitational lenses deflect light so strongly, or why visible matter is distributed as it is both in the local universe and in cosmic space. However, it is not known whether dark matter is formed by undiscovered particles or whether it can be explained by modifying the standard laws of physics. It is in this scenario that advances in the discovery of new solutions in the framework of 5D gravitational theories can provide satisfactory explanations for these astronomical measurements at present.

\section*{Data aviability statement} 

This manuscript has no associated data.  
Data sharing
not applicable to this article as no datasets were generated or analysed
during the current study.

\section*{Acknowledgments}
This  work  was  partially supported by the 
Grant PID2021-122938NB-I00 funded by MCIN/AEI/
10.13039/501100011033 and by “ERDF A way of making Europe”, as well as  the Consejer\'\i a
de Educaci\'on of the Junta de Castilla y Le\'on under the Research Project Grupo
de Excelencia GR234 Ref.:SA096P20 (Fondos Feder y
en l\'\i nea con objetivos RIS3).


\end{document}